\documentclass[preprint,showpacs,preprintnumbers,amsmath,amssymb]{revtex4}

\usepackage{graphicx}
\usepackage{dcolumn}
\usepackage{bm}

\begin{document}
\preprint{To appear in Physical Plasmonics (2006)}

\title{Surface Plasmon Polariton Mach-Zehnder Interferometer and Oscillation Fringes}

\author{A.~Drezet, A.~Hohenau, A.~L.~Stepanov, H.~Ditlbacher,
B.~Steinberger, F.~R.~Aussenegg, A.~Leitner, and J.~R.~Krenn}
\affiliation{Institute of physics,
Karl-Franzens University Graz, Universit\"atsplatz 5 A-8010 Graz,
Austria}
\date{\today}

\begin{abstract}
We present a quantitative experimental analysis of a surface plasmon
polariton (SPP) interferometer relying on elliptical Bragg mirrors. By using a
leakage radiation microscope we observe oscillation fringes with unit
visibility at the two interferometer exits. We study the properties of the SPP
beam splitter and determine experimentally both the norm and phase of the SPP
reflection and transmission coefficients.
\end{abstract}

 \maketitle

Progress in the field of two-dimensional optics at the micro- and nanometer
scale requires necessarily the control over coherence of wave propagation in a
confined environment. In this context it has been experimentally shown a few
years ago that surface plasmon polaritons (SPPs, electromagnetic waves
confined at the interface of a metal and a dielectric \cite{Raether}) can
generate a variety of two-dimensional interference effects
\cite{Hecht,Sergy,Harry2}.

More recently we were able to implement two-dimensional SPP Mach-Zehnder
interferometers fabricated on silver thin films by lithographic techniques
\cite{Harry1,Aurelien1}. The observation of SPP propagation relied on the
interaction of a layer of fluorescent molecules deposited on top of the SPP
sustaining silver surface. The fluorescence intensity was detected with
conventional far-field microscopy providing a map of the lateral SPP field
intensity profile \cite{Harry2}. As a major drawback, fluorescence imaging
could not be exploited quantitatively as the dye molecules photo-bleach
rapidly, and do so in dependence of the local SPP intensity.

For this reason we shifted our attention towards leakage radiation microscopy
(LRM) which relies on the coupling of SPP waves to leaky light modes
propagating in the dielectric substrate supporting the metal thin film
\cite{Hecht,Bouhelier}. The leakage radiation is collected by an immersion
objective optically coupled to the substrate which allows to acquire an image
of the SPP profile in the metal surface \cite{Andrey}. The advantage of LRM
is that quantitative analysis can be performed since the intensity recorded at
any point of the microscope image plane is proportional to the SPP intensity
at the conjugate point in the object plane. We used LRM previously to analyze
the interaction of SPPs with a line of gold protrusions on a silver film
acting as a SPP beam splitter \cite{Andrey}, as well as the propagation of
SPPs in a corral constituted by confocal elliptical Bragg mirrors on a gold
film \cite{Aurelien1}.

Here, we present a quantitative analysis of a Mach-Zehnder interferometer for
SPP waves and we show in particular that we can record interference
oscillation with unit visibility at the two interferometer exits.

\begin{figure}[h]
\includegraphics[angle=-90,width=3in]{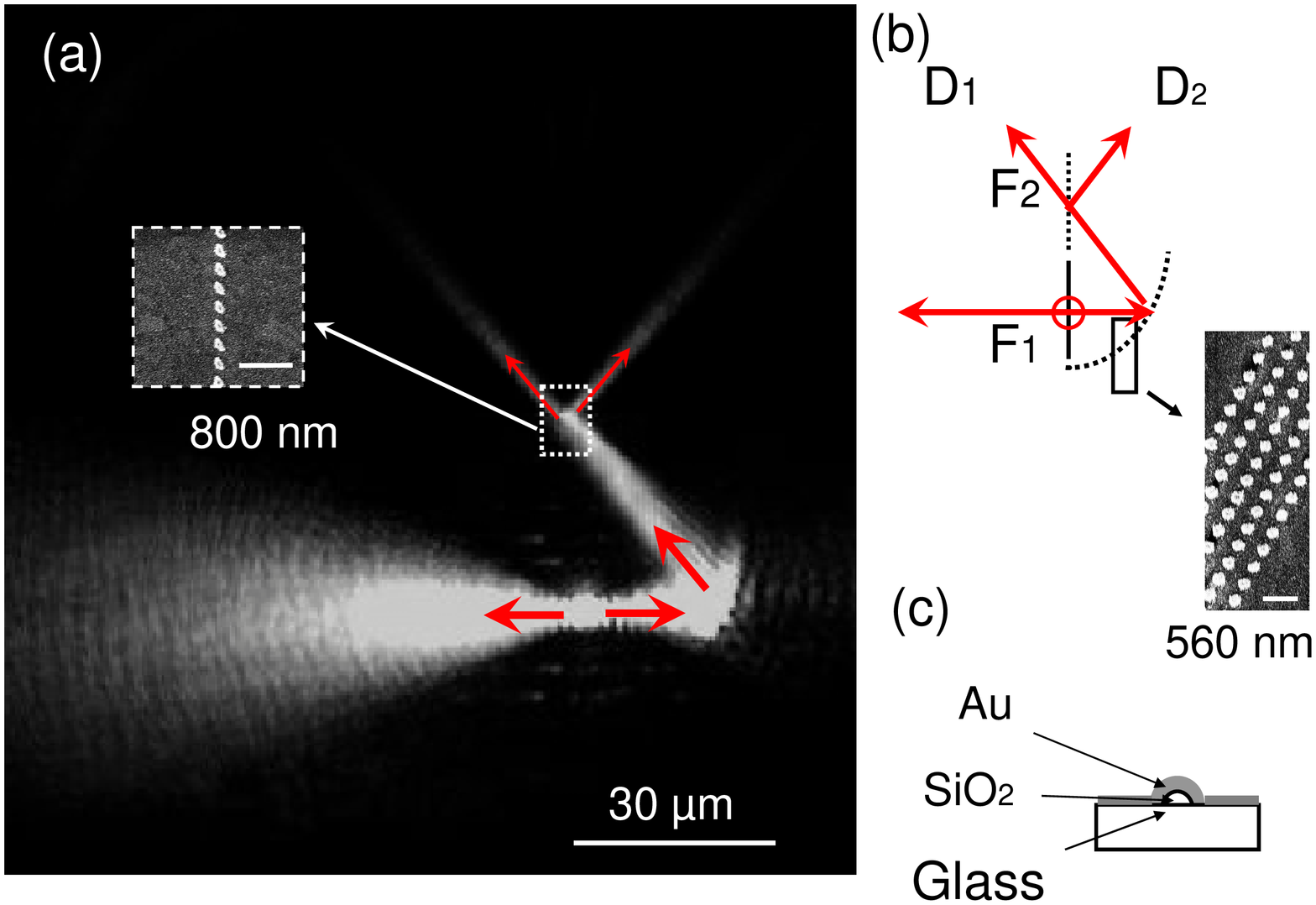}
\caption{(color online) (a) LRM image of the half-interferometer
as described in the text. The inset shows a scanning electron
microscope image of the SPP beam splitter. (b) Sketch of the
half-interferometer. The two reflected SPP beams, originating from
the ridge at $F_{1}$, intersect at the second focal point $F_{2}$
of the elliptical mirror where the beam splitter $BS$ is located.
The inset shows a scanning electron microscope image of the Bragg
mirror. (c) Sketch of an individual protrusion building the beam
splitter and the Bragg mirrors.} \label{FigA}
\end{figure}

The experimental basics of the SPP interferometer considered here
rely on results obtained in Refs.~\cite{Harry1} and
\cite{Aurelien1}. SPPs are launched locally on a gold thin film by
focussing a laser beam (Titane:Sapphire, $\lambda_{0}=750$ nm)
incident normally to the substrate through a microscope objective
(50$\times$, numerical aperture $0.7$) on a gold ridge ($160$ nm
width, $80$ nm height) on the film. The ridge and all other
structures on the gold film as discussed in the following were
fabricated by electron beam lithography (EBL). First, the
structure geometries are defined by electron beam exposure,
various chemical development steps and the deposition of $80$ nm
thick SiO$_{2}$. Then, second, the whole substrate is covered by a
$80$ nm thick gold film. The SPPs propagate in the directions
normal to the ridge axis and are reflected upon interaction with
elliptical Bragg mirrors. The Bragg mirrors are made of individual
protrusions of $185$ nm diameter and $280$ nm center-to-center
distance. These protrusions are arranged in order to constitute
five confocal ellipses. The distance between the two ellipse foci
$F_{1}$ and $F_{2}$ equals the minimal long axis length
$a_{min}=30 $ $\mu m$. Since we work with a Bragg mirror the
variation $\delta a$ of the long axis length between two
consecutive ellipses must be $N\lambda_{SPP}/2$ where $N$ is an
integer and $\lambda_{SPP}\simeq\lambda_{0}$ is the SPP wavelength
fixed here at $\lambda_{SPP} \approx 750 $ nm. We choose $N=1$ and
$\delta a=375$ nm. With these parameters we achieve mirror
reflectivity up to $90\%$. As in Ref.~\cite{Harry1} we consider a
Mach-Zehnder configuration. The two SPP beams launched at the
ridge and reflected by the elliptical Bragg mirror are focused
onto a beam splitter made of a line of individual protrusions
($160$ nm diameter, center-to-center distance $240$ nm). In order
to exploit the optical properties of elliptical mirrors the SPPs
are launched at the first focal point $F_{1}$. Accordingly, the
beam splitter $BS$ is positioned at the second focal point
$F_{2}$, where the two reflected SPP beams intersect (see
Fig.~\ref{FigA} and Fig.~\ref{FigD}). As we will see later we can
adjust the phase in the interferometer by changing the lateral
position of the ridge within different interferometers. Therefore,
we consider several similar interferometers built by EBL on the
same sample.

\begin{figure}[h]
\includegraphics[width=3in]{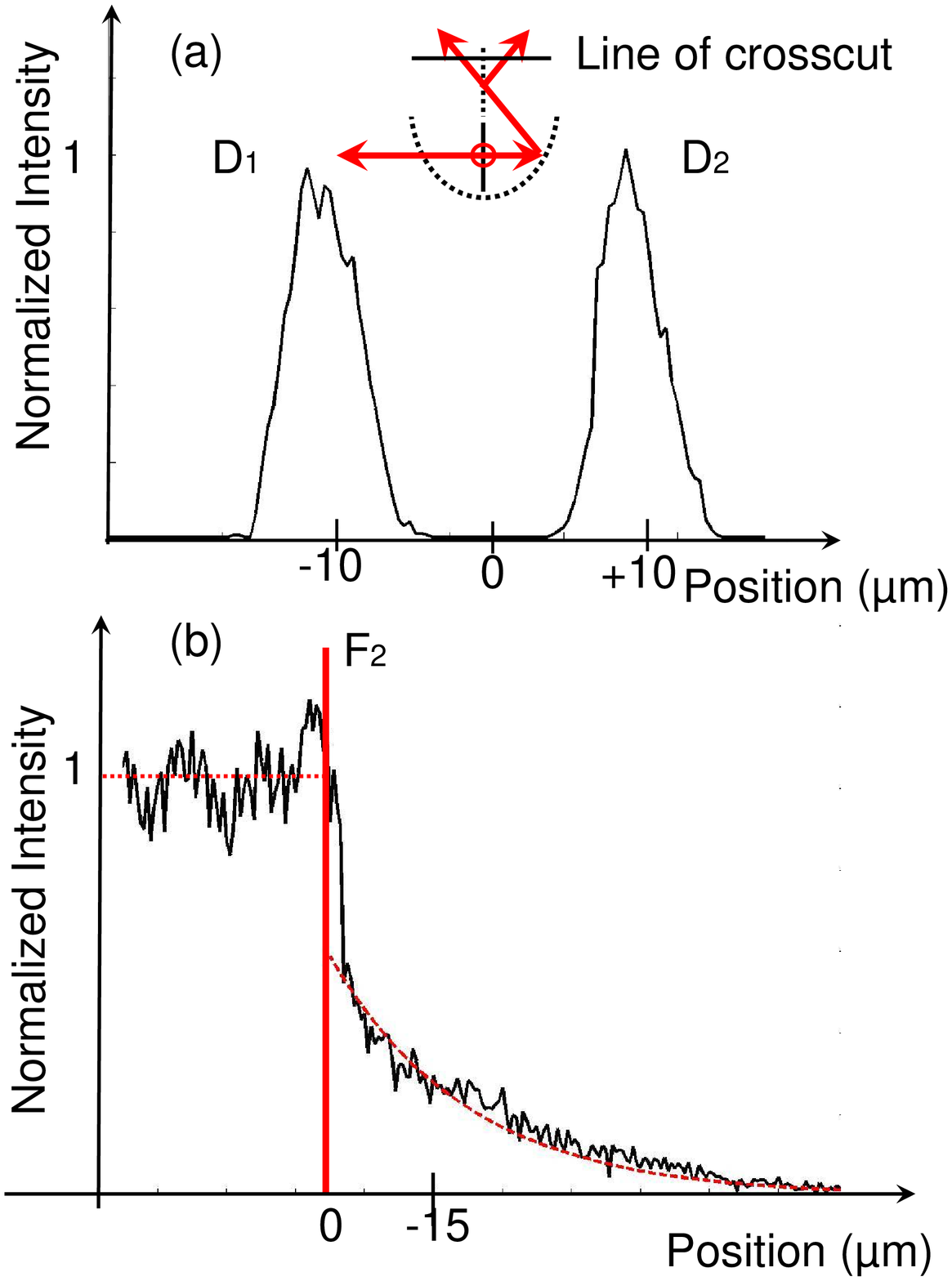}
\caption{(color online) (a) Transversal cross-cut of the two exit
SPP beams, see inset. (b) Longitudinal cross-cut of the SPP beam
going through $F_{2}$ (indicated by the vertical red line) and
$D_{1}$. The red dotted curve is an exponential fit and the
horizontal red dotted line is centered on the average SPP
intensity value in the left part of the curve.} \label{FigB}
\end{figure}

We start our investigation by considering a half-interferometer,
i.e., a configuration where only one half of the elliptical Bragg
mirror is present (see Fig.~\ref{FigA}). SPP waves propagating
into the left arm are thus not reflected and cannot interfere with
the SPP waves propagating in the right arm. As a consequence the
SPP wave initially launched to the right splits at $BS$ into two
output SPP waves $D_{1}$ and $D_{2}$ with normalized intensity
$I_{1}=|T|^{2}$ (transmissivity), and $I_{2}=|R|^{2}$
(reflectivity) such that $|R|^{2}+|T|^{2}=1$ assuming no
scattering. Fig.~1a shows a LRM image corresponding to this
half-interferometer configuration and revealing a perfect symmetry
between $D_{1}$ and $D_{2}$.  Fig.~\ref{FigB}a shows a transverse
cross-cut of the intensity in the two output SPP beams which
confirms this symmetry quantitatively.

\begin{figure}[h]
\includegraphics[width=3in]{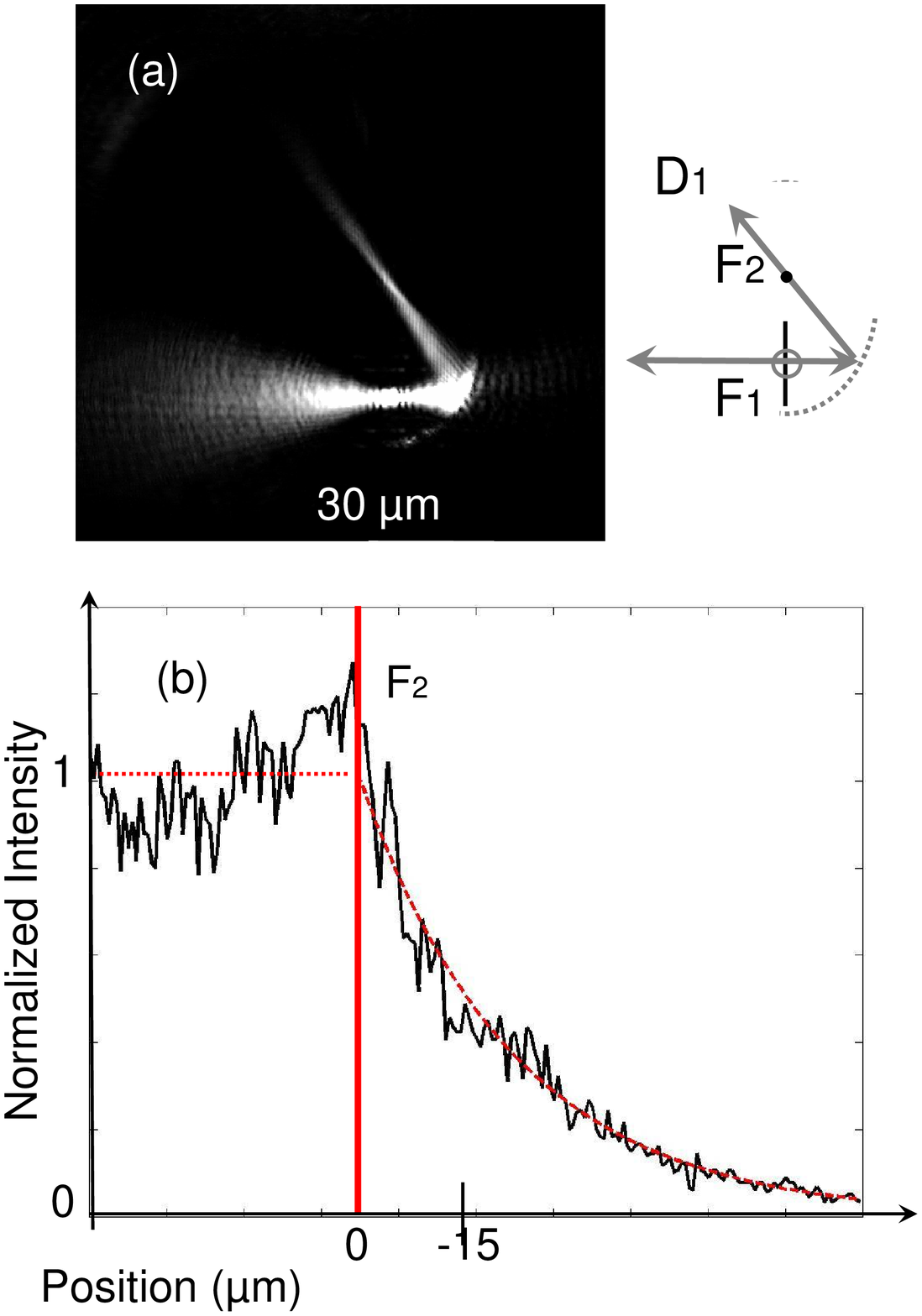}
\caption{(color online) (a) LRM image of a configuration without
SPP beam splitter (see inset). (b) Longitudinal SPP intensity
cross-cut from (a), corresponding to Fig.~2b.} \label{FigC}
\end{figure}

\begin{figure}[h]
\includegraphics[width=4in]{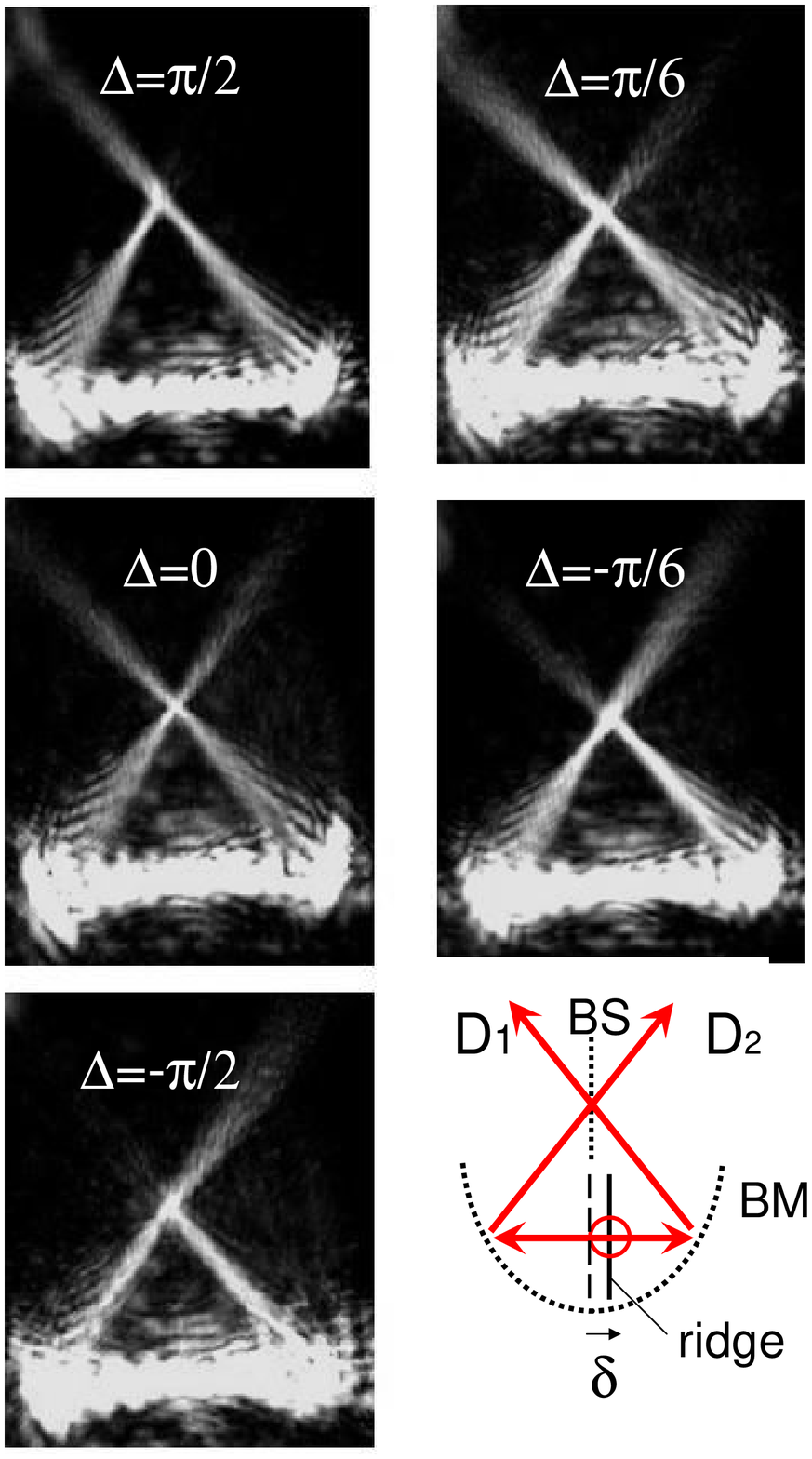}
\caption{(color online) Sequence of LRM images of SPP propagation
in Mach-Zehnder interferometers. The interferometer configurations
correspond to phase differences of $\Delta=-\pi/2,-\pi/6,0,\pi/6$,
and $\pi/2$, respectively. The inset shows how the phase
difference is introduced by displacing the SPP launching ridge by
an (algebraic) amount $\delta$. $D_{1}$ and $D_{2}$ are the two
exit arms and $BS$ is the beam splitter. The two reflected SPP
beams intersect at the second focal point of the elliptical Bragg
mirror $BM$ at $BS$.} \label{FigD}
\end{figure}

Fig.~\ref{FigB}b shows a cross-cut along the SPP beam reflected by
the right Bragg mirror, i.e., through $F_{2}$ and $D_{1}$.
Comparison with the corresponding cross-cut for a sample where no
beam splitter $BS$ is present (see Fig.~\ref{FigC}) shows that the
SPP intensity is reduced by a factor of two by the presence of the
beam splitter. This justifies the assumption that no scattering
takes place. The presented results are all consistent with the
parameters of a 50/50 lossless beam splitter,
i.~e.~$|R|^{2}\simeq|T|^{2}\simeq 1/2$. The SPP intensity decay in
the branch $D_{1}$ located after $F_{2}$ can be well reproduced by
an exponential function $I_{\textrm{transmitted}}(r)\propto
e^{-r/L_{SPP}}$ where $r$ is the distance separating the
observation point from $F_{2}$, and $L_{SPP}\simeq 20$ $\mu$m is
the SPP propagation length. The latter value agrees well with
literature values for a $80$ nm thick gold film \cite{Raether}.

It is important to note that with such an half-interferometer we do not have
experimental access to the phase relation between the transmitted, reflected,
and incident beams. In order to define this relation we must first recall some
properties of a lossless $BS$. From the unitarity of the $BS$ transfer matrix
\cite{Zeilinger} and from energy conservation one can deduce that the relation
between the reflection and transmission amplitudes $R$ and $T$ is such that
$T=Re^{i\phi}$ where $\phi=\phi_{\pm}=\pm \pi/2$. However, the sign $\pm 1$
depends on internal properties of the beam splitter and cannot be deduced from
simple unitarity considerations. An individual study of each physical case is
thus necessary. The simplest experimental way to do that is to consider the
interferometer configuration including both (left and right) parts of the
elliptical Bragg mirror, i.~e.~the complete interferometer. However, in order
to explain the predictions of this interferometry experiment one has to take
carefully into consideration all the phase shifts and differences introduced
during SPP propagation.

In this context it has been remarked \cite{Theseharry} that one
can mimic the behavior of SPP waves launched from $F_{1}$ on the
ridge by using the scalar field produced by a linear and
continuous distribution of 2D dipoles. These in-plane dipoles are
orthogonal to the ridge and proportional in strength to the
incident electric field \cite{Remark2}. From this hypothesis and
from symmetry considerations we deduce that there is a phase
difference of $\pi$ between the SPPs propagating into the right
and into the left arm of the interferometer. The pertinence of
this fact has been experimentally observed for the elliptical SPP
corral \cite{Aurelien1}. Indeed, this phase difference justifies
why we can observe SPP intensity oscillations at the second focal
point $F_{2}$ when rotating the in-plane laser beam polarization
at $F_{1}$. This result is not in conflict with the previous
analysis of the symmetric and lossless beam splitter. Indeed, the
phase difference of $\pm \pi/2$ characterizes an ideal lossless
beam splitter coupling an incident mode to two output modes.

However, the beam splitter is supposed to be lossless, as confirmed by our
first analysis of the half-interferometer case, and we thus deduce that the
intensities in the two exits $D_{1}$ and $D_{2}$ of the interferometer are
given by
\begin{eqnarray}
I_{1}=\frac{1}{2}|\alpha|^{2}|Re^{ik_{SPP}\delta}-Te^{-ik_{SPP}\delta}|^{2}
\nonumber\\
I_{2}=\frac{1}{2}|\alpha|^{2}|Te^{ik_{SPP}\delta}-Re^{-ik_{SPP}\delta}|^{2}.
\end{eqnarray}
$|\alpha|^{2}$ is the SPP coupling efficiency at the ridge.
$\delta$ is the algebraic displacement of the ridge with respect
to the symmetry axis of the interferometer (defining the symmetric
configuration) and $k_{SPP}=2\pi/\lambda_{SPP}$ is the real part
of the SPP wave vector. $\delta$ defines a variable phase
difference responsible for the oscillation of the intensities at
$D_{1}$ and $D_{2}$. We note that we neglected the small
contribution of the imaginary part of the SPP wave vector
$1/(2L_{SPP})$ since $e^{\pm\delta/L_{SPP}}\simeq 1$ for
$\delta\ll L_{SPP}$ \cite{Remark}. Depending on the sign of the
phase shift $\phi_{\pm}=\pm\pi/2$ at the beam splitter we obtain
two possible solutions
\begin{eqnarray}
I_{1,\pm}=\frac{1}{2}|\alpha|^{2}(1\pm2|R||T|\sin{(2k_{SPP}\delta)})
\nonumber\\
I_{2,\pm}=\frac{1}{2}|\alpha|^{2}(1\mp2|R||T|\sin{(2k_{SPP}\delta)}),
\end{eqnarray} which invert the role of $D_{1}$ and
$D_{2}$, respectively.

\begin{figure}[h]
\includegraphics[angle=-90,width=3.5in]{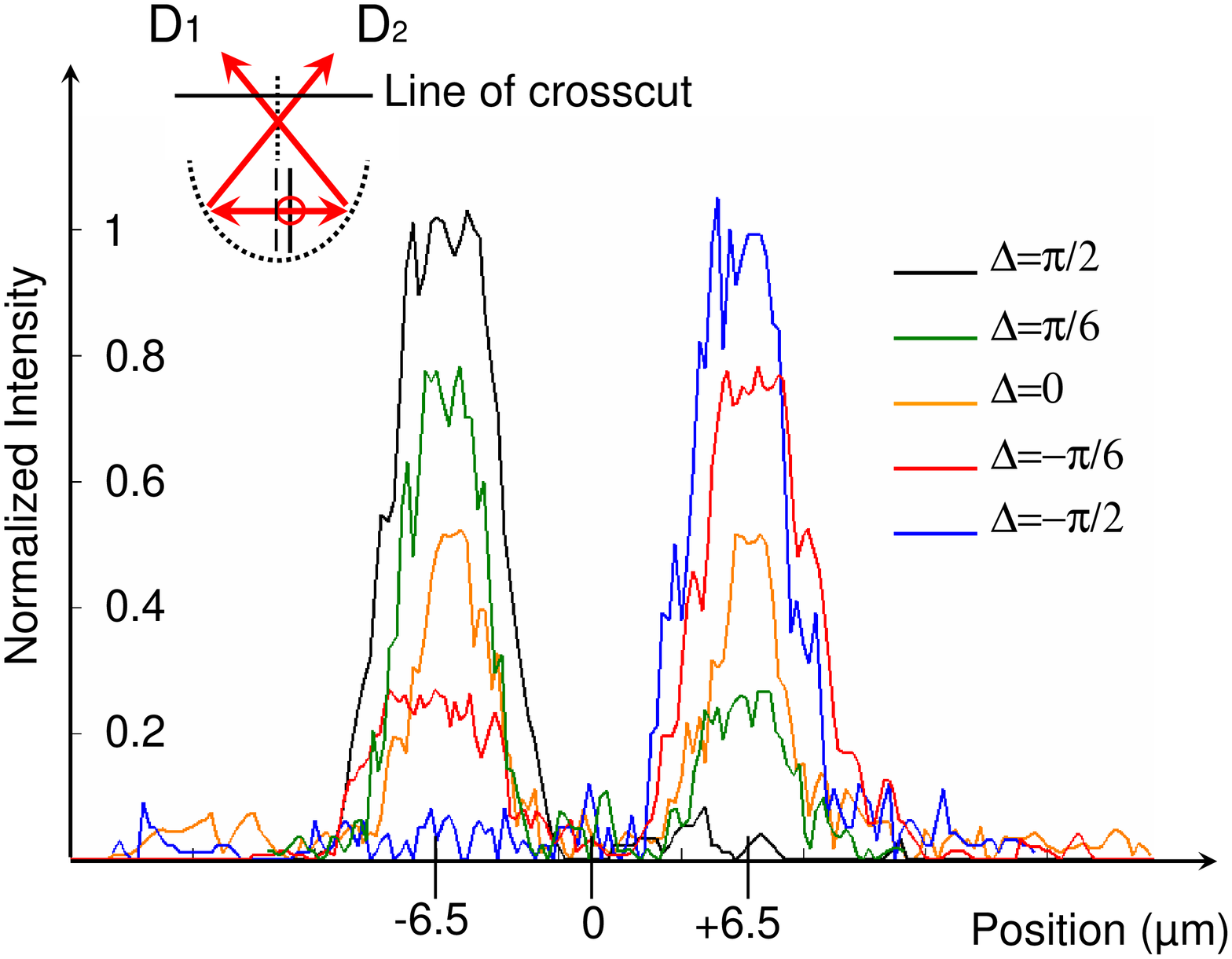}
\caption{(color online) Sequence of cross-cuts of the SPP
intensity measured in Fig.~\ref{FigF}. The position of the
cross-cut direction is indicated in the inset by the horizontal
line intersecting the two beams $D_{1}$, and $D_{2}$. This line is
located at 11 $\mu$m above the second focal point $F_{2}$. The
according phase shifts are given in the figure legend.}
\label{FigE}
\end{figure}

\begin{figure}[h]
\includegraphics[angle=-90,width=3.5in]{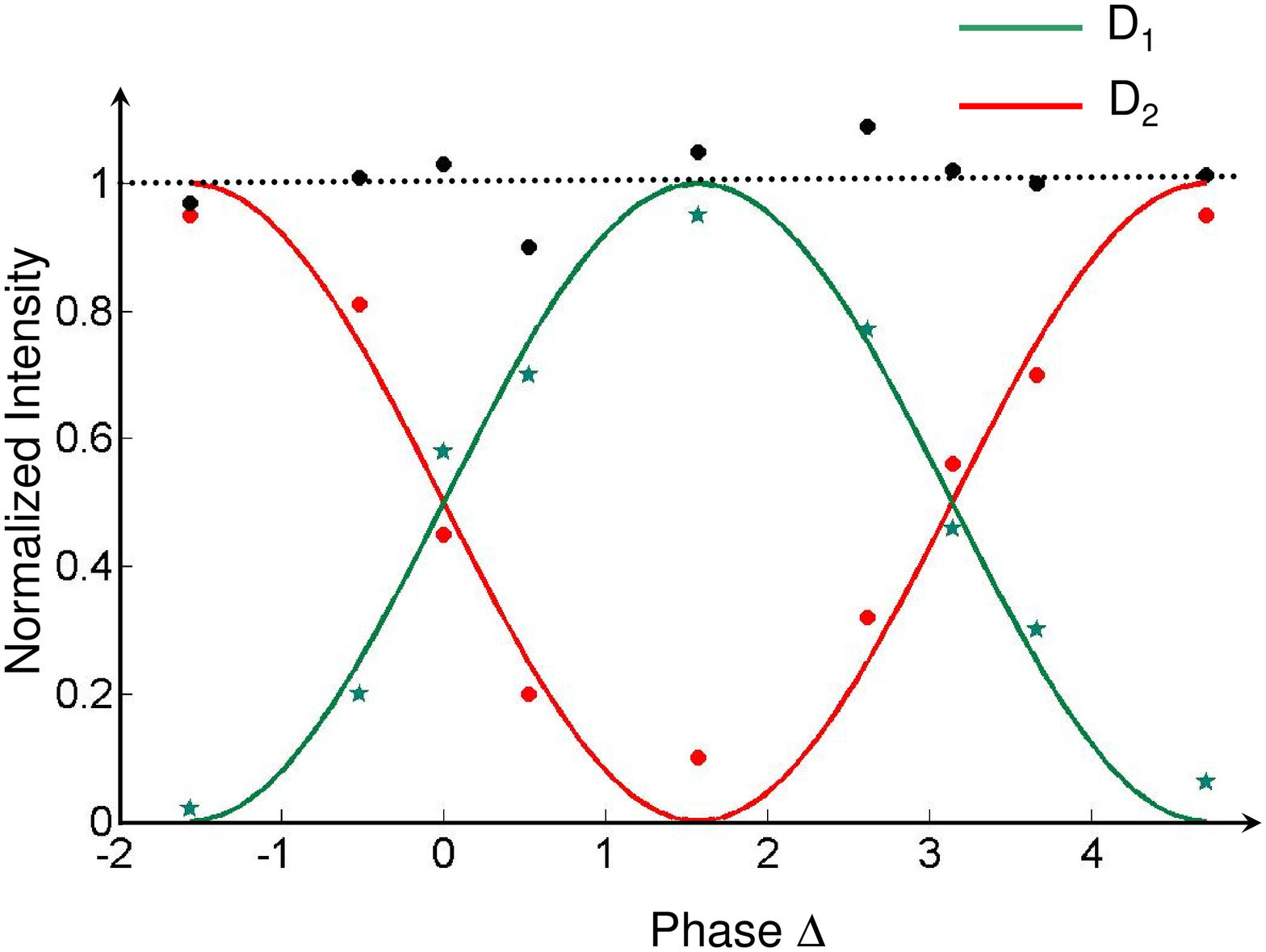}
\caption{(color online) Interferogram of the two exits $D_{1}$ and
$D_{2}$, intensity normalized to maximum values. The experimental
points used (stars for $I_{1}$, circles for $I_{2}$, squares for
the sum of both $I_{1}$ and $I_{2}$) are obtained from cross-cuts
similar to the one represented in Fig.~\ref{FigE}. The green and
red lines represent the theoretically expected interferograms at
$D_{1}$ and $D_{2}$, respectively. They are obtained from the
formula Eq.~2 in the case of a $BS$ with $T=i/\sqrt{2}$ and
$R=1/\sqrt{2}$. The dotted horizontal line represents the energy
conservation rule $|R|^{2}+|T|^{2}=1$.} \label{FigF}
\end{figure}

Experimentally we changed discontinuously $\delta$ between
different individual interferometers fabricated on one sample to
obtain a variation of the phase $\Delta=2k_{SPP}\delta$ in the
domain $\Delta \in[-\pi/2,\pi/2]$. Fig.~\ref{FigD} shows the
according sequence of LRM images corresponding to phase
differences $\Delta=-\pi/2,-\pi/6,0,\pi/6$, and $\pi/2$. Since the
SPP wavelength $\lambda_{SPP}$ is fixed to $750$ nm the
displacement $\delta$ is thus varying in the interval
$[-\lambda_{SPP}/8,+\lambda_{SPP}/8]\simeq [-94\textrm{ nm},+94
\textrm{nm}]$. In this sequence of images we clearly observe the
intensity oscillation in the SPP beams $D_{1}$ and $D_{2}$ as a
function of $\Delta$. In order to be quantitative we consider
transversal cross-cuts of the two exit beams, corresponding to the
cross-cut in Fig.~\ref{FigB}. All these cross-cuts were taken from
the same position for all interferometers,i.~e.~for all values of
$\Delta$ (see inset on Fig.~\ref{FigE}). From this analysis one
can deduce the interferograms plotted in Fig.~\ref{FigF}.

Clearly the two plots corresponding to the exits $D_{1}$ and
$D_{2}$ are in opposition of phase as requested for a Mach-Zehnder
interferometer. Energy conservation is fulfilled since
$I_{1}+I_{2}\simeq 1$ after normalization (see Fig.~\ref{FigF}).
The data show a perfect consistency with the model given by Eq.~1
in the case of a $BS$ phase shift $\phi_{+}=+\pi/2$. The fringe
visibility \cite{Born}
$V=(I_{\textrm{max}}-I_{\textrm{min}})/(I_{\textrm{max}}+I_{\textrm{min}})$
is close to unity and agrees with the values $|R|^{2}\simeq
|T|^{2}\simeq 1/2$ deduced from the expression
$V=2|R||T|/(|R|^{2}+|T|^{2})$ (submitted to the energy
conservation condition $|R|^{2}+|T|^{2}=1$). We can thus with a
good approximation write $T=i/\sqrt{2}$ and $R=1/\sqrt{2}$.

To conclude, we presented a detailed experimental analysis of SPP
interferometry. Therefore, we analyzed LRM images of a Mach-Zehnder
interferometer relying on elliptical Bragg mirrors, and we analyzed
quantitatively the reflection and transmission of SPP waves by the lossless
beam splitter $BS$. This leads to the experimental relations $T=i/\sqrt{2}$
and $R=1/\sqrt{2}$ between the SPP transmission and reflection coefficients
$T,R$ of $BS$. We observed fringes oscillations with unit visibility at the
two exits of the interferometer as well as the opposition of phase between the
two signals $I_{1}$ and $I_{2}$ when changing the length of the two
interferometer arms by an amount $\pm \delta$, i.~e.~when changing the phase
by the quantity $\Delta=2k_{SPP}\delta$. The conservation of energy is
fulfilled since we have $I_{1}+I_{2}\simeq const.$ whatever the phase
difference $\Delta$. All experimental observations are in good quantitative
agreement with the analytical calculations. This forms a solid base for
further investigations or applications of SPP interferometry.

For financial support the Austrian Science Foundation and the European Union,
under projects FP6 NMP4-CT-2003-505699 and FP6 2002-IST-1-507879 are
acknowledged.


\begin{references}

\bibitem{Raether}
H.~Raether, \emph{Surface Plasmons}(Springer, Berlin, 1988).
\bibitem{Hecht}
B.~Hecht, H.~Bielefeldt, L.~Novotny, Y.~Inouye, and D.~W.~Pohl,
Phys.~Rev.~Lett.~\textbf{77}, 1889 (1996).
\bibitem{Sergy}
S.~I.~Bozhevolnyi, and F.~A.~Pudonin,
Phys.~Rev.~Lett.~\textbf{78}, 2823 (1997).
\bibitem{Harry2}
H.~Ditlbacher, J.~R.~Krenn, N.~Felidj, B.~Lamprecht, G.~Schider,
M.~Salemo, A.~Leitner, and F.~R.~Aussenegg,
Appl.~Phys.~Lett.~\textbf{80}, 404 (2002).
\bibitem{Harry1}
H.~Ditlbacher, J.~R.~Krenn, G.~Schider, A.~Leitner, and
F.~R.~Aussenegg, Appl.~Phys.~Lett.~\textbf{81}, 1762 (2002).
\bibitem{Aurelien1}
A.~Drezet, A.~L.~Stepanov, H.~Ditlbacher, A.~Hohenau,
B.~Steinberger, F.~R.~Aussenegg, A.~Leitner, and J.~R.~Krenn,
Appl.~Phys.~Lett.~\textbf{86}, 074104 (2005).
\bibitem{Bouhelier}
A.~Bouhelier, Th.~Huser, H.~Tamaru, H.~-J.~G\"{u}ntherodt,
D.~W.~Pohl, Fadi I.~Baida and D.~Van Labeke,
Phys.~Rev.~B.~\textbf{63}, 155404 (2001).
\bibitem{Andrey}
A.~stepanov, J.~R.~Krenn, H.~Ditlbacher, A.~Hohenau, A.~Drezet,
B.~Steinberger, A.~Leitner, and F.~Aussenegg,
Opt.~Lett.~\textbf{30},1524 (2005)
\bibitem{Zeilinger}
A.~Zeilinger, Am.~J.~Phys.~\textbf{49}, 882 (1981).\\
C.~H.~Holbrow, E.~Galvez, and M.~E.~Parks,
Am.~J.~Phys.~\textbf{70}, 260 (2002).
\bibitem{Theseharry}
H.~Ditlbacher, Phd Thesis, Karl-Franzens Universit\"{a}t Graz
(2003).
\bibitem{Remark2}
Each dipole generate at $x,y$  a scalar field
\begin{eqnarray*}\Psi(x,y)\propto
\cos{\left(\Theta\right)}e^{ik_{SPP}r}e^{-r/(2L_{SPP})}/\sqrt{r}\end{eqnarray*}
\cite{Hecht,Harry2,Drezet}. $r=\sqrt{(x-x_{s})^{2}+(y-y_{s})^{2}}$
is the distance separating the source located at $x_{s},y_{s}$
from  the observation point, and $\Theta$ is the angle between the
dipole $\mathbf{P}$ and the vector position $\mathbf{r}$ going
from the source to the observation point. The dipole is induced by
the local electric field $\mathbf{E}_{0}(x_{s},y_{s})$ and thus
$\mathbf{P}=\chi\mathbf{E}_{0}(x_{s},y_{s})$ where $\chi$ is the
dipole polarizability.
\bibitem{Drezet} M.~Brun, A.~Drezet,
H.~Mariette, N.~Chevalier, J.~C.~Woehl, and S.~Huant, Europhys.
Lett.~\textbf{64}, 634 (2003).
\bibitem{Remark} If we don't neglect the
SPP damping we obtain
$V_{1}=2|R||T|/(|R|^{2}e^{-\delta/(2L_{SPP})}+|T|^{2}e^{-\delta/(2L_{SPP})})$
and
$V_{2}=2|R||T|/(|R|^{2}e^{-\delta/(2L_{SPP})}+|T|^{2}e^{-\delta/(2L_{SPP})})$
for respectivelly the signal $D_{1}$ and $D_{2}$. Here we have
$V_{1}\simeq V_{2}=V$.
\bibitem{Born} M.~Born and E.~Wolf, \emph{Principles of Optics}
(Cambridge University Press, Cambridge, 1999).
\end{references}
\end{document}